\def\cerntp{1}

\def\opus{1}

\documentstyle[12pt,epsfig,ifthen,citesort]{article}

\newcommand{\ifdraft}{nodraft}

\newcommand{\ifpsdraft}{nodraft}

\newcommand{\vdate}{July 1996}
\newcommand{\cernnr}{96--187}
\newcommand{\beq}{\begin{equation}}
\newcommand{\eeq}{\end{equation}}
\newcommand{\beqn}{\begin{eqnarray}}
\newcommand{\eeqn}{\end{eqnarray}}
\newcommand{\epem}{$\mbox{\it e}^+ \mbox{\it e}^-$}

\newcommand{\GeV}{\mbox{GeV}}

\newcommand{\dd}{\mbox{d}}

\newcommand{\fullline}{
\unitlength0.4mm
\begin{picture}(13,4)
\linethickness{0.3mm}
\put(-1,2.0){\line(1,0){15}}
\thinlines
\end{picture}
}

\newcommand{\dashline}{
\unitlength0.4mm
\begin{picture}(20,4)
\linethickness{0.3mm}
\put(-1,2.0){\line(1,0){4}}
\put(8,2.0){\line(1,0){4}}
\put(17,2.0){\line(1,0){4}}
\thinlines
\end{picture}
}

\newcommand{\dgpicture}[2]{
\begin{picture}(#1,#2)
\thicklines
\thinlines
}

\newcommand{\epsfigdg}[2]{\epsfig{figure=#1,#2}}

\newcommand{\chsign}[1]{%
{\ifthenelse{\equal{\ifdraft}{draft}}%
{
{\sf {\Large$\bullet$$\bullet$$\bullet$} #1 
                         {\Large$\bullet$$\bullet$$\bullet$} }%
}%
{}%
}}
\newcommand{\smallmark}[1]{
\marginpar{\fbox{\vspace{0.0cm}{\scriptsize #1}}}}

\newcommand{\labelm}[1]{%
\label{#1}%
\ifthenelse{\equal{\ifdraft}{draft}}%
{\smallmark{#1}}%
{}%
}

\newcommand{\labelmm}[1]{%
\label{#1}%
\ifthenelse{\equal{\ifdraft}{draft}}%
{\protect\fbox{\sf #1}}%
{}%
}

\newcommand{\beqm}[1]{%
\ifthenelse{\equal{\ifdraft}{draft}}%
{\smallmark{#1}}%
{}%
\beq \label{#1}}

\newcommand{\beqnm}[1]{%
\ifthenelse{\equal{\ifdraft}{draft}}%
{\smallmark{#1}}%
{}%
\beqn \label{#1}}

\ifthenelse{\equal{\ifpsdraft}{draft}}{\psdraft}{}

\catcode`\@=11

 \font\tenmsx=msam10 scaled \magstep1
 \font\sevenmsx=msam8
 \font\fivemsx=msam6
 \font\tenmsy=msbm10 scaled \magstep1
 \font\sevenmsy=msbm8
 \font\fivemsy=msbm6

\newfam\msxfam
\newfam\msyfam
\textfont\msxfam=\tenmsx  \scriptfont\msxfam=\sevenmsx
  \scriptscriptfont\msxfam=\fivemsx
\textfont\msyfam=\tenmsy  \scriptfont\msyfam=\sevenmsy
  \scriptscriptfont\msyfam=\fivemsy

\def\hexnumber@#1{\ifnum#1<10 \number#1\else
 \ifnum#1=10 A\else\ifnum#1=11 B\else\ifnum#1=12 C\else
 \ifnum#1=13 D\else\ifnum#1=14 E\else\ifnum#1=15 F\fi\fi\fi\fi\fi\fi\fi}

\def\msx@{\hexnumber@\msxfam}
\def\msy@{\hexnumber@\msyfam}
\mathchardef\boxdot="2\msx@00
\mathchardef\boxplus="2\msx@01
\mathchardef\boxtimes="2\msx@02
\mathchardef\square="0\msx@03
\mathchardef\blacksquare="0\msx@04
\mathchardef\centerdot="2\msx@05
\mathchardef\lozenge="0\msx@06
\mathchardef\blacklozenge="0\msx@07
\mathchardef\circlearrowright="3\msx@08
\mathchardef\circlearrowleft="3\msx@09
\mathchardef\rightleftharpoons="3\msx@0A
\mathchardef\leftrightharpoons="3\msx@0B
\mathchardef\boxminus="2\msx@0C
\mathchardef\Vdash="3\msx@0D
\mathchardef\Vvdash="3\msx@0E
\mathchardef\vDash="3\msx@0F
\mathchardef\twoheadrightarrow="3\msx@10
\mathchardef\twoheadleftarrow="3\msx@11
\mathchardef\leftleftarrows="3\msx@12
\mathchardef\rightrightarrows="3\msx@13
\mathchardef\upuparrows="3\msx@14
\mathchardef\downdownarrows="3\msx@15
\mathchardef\upharpoonright="3\msx@16

\mathchardef\downharpoonright="3\msx@17
\mathchardef\upharpoonleft="3\msx@18
\mathchardef\downharpoonleft="3\msx@19
\mathchardef\rightarrowtail="3\msx@1A
\mathchardef\leftarrowtail="3\msx@1B
\mathchardef\leftrightarrows="3\msx@1C
\mathchardef\rightleftarrows="3\msx@1D
\mathchardef\Lsh="3\msx@1E
\mathchardef\Rsh="3\msx@1F
\mathchardef\rightsquigarrow="3\msx@20
\mathchardef\leftrightsquigarrow="3\msx@21
\mathchardef\looparrowleft="3\msx@22
\mathchardef\looparrowright="3\msx@23
\mathchardef\circeq="3\msx@24
\mathchardef\succsim="3\msx@25
\mathchardef\gtrsim="3\msx@26
\mathchardef\gtrapprox="3\msx@27
\mathchardef\multimap="3\msx@28
\mathchardef\therefore="3\msx@29
\mathchardef\because="3\msx@2A
\mathchardef\doteqdot="3\msx@2B

\mathchardef\triangleq="3\msx@2C
\mathchardef\precsim="3\msx@2D
\mathchardef\lesssim="3\msx@2E
\mathchardef\lessapprox="3\msx@2F
\mathchardef\eqslantless="3\msx@30
\mathchardef\eqslantgtr="3\msx@31
\mathchardef\curlyeqprec="3\msx@32
\mathchardef\curlyeqsucc="3\msx@33
\mathchardef\preccurlyeq="3\msx@34
\mathchardef\leqq="3\msx@35
\mathchardef\leqslant="3\msx@36
\mathchardef\lessgtr="3\msx@37
\mathchardef\backprime="0\msx@38
\mathchardef\risingdotseq="3\msx@3A
\mathchardef\fallingdotseq="3\msx@3B
\mathchardef\succcurlyeq="3\msx@3C
\mathchardef\geqq="3\msx@3D
\mathchardef\geqslant="3\msx@3E
\mathchardef\gtrless="3\msx@3F
\mathchardef\sqsubset="3\msx@40
\mathchardef\sqsupset="3\msx@41
\mathchardef\vartriangleright="3\msx@42
\mathchardef\vartriangleleft="3\msx@43
\mathchardef\trianglerighteq="3\msx@44
\mathchardef\trianglelefteq="3\msx@45
\mathchardef\bigstar="0\msx@46
\mathchardef\between="3\msx@47
\mathchardef\blacktriangledown="0\msx@48
\mathchardef\blacktriangleright="3\msx@49
\mathchardef\blacktriangleleft="3\msx@4A
\mathchardef\vartriangle="3\msx@4D
\mathchardef\blacktriangle="0\msx@4E
\mathchardef\triangledown="0\msx@4F
\mathchardef\eqcirc="3\msx@50
\mathchardef\lesseqgtr="3\msx@51
\mathchardef\gtreqless="3\msx@52
\mathchardef\lesseqqgtr="3\msx@53
\mathchardef\gtreqqless="3\msx@54
\mathchardef\Rrightarrow="3\msx@56
\mathchardef\Lleftarrow="3\msx@57
\mathchardef\veebar="2\msx@59
\mathchardef\barwedge="2\msx@5A
\mathchardef\doublebarwedge="2\msx@5B
\mathchardef\angle="0\msx@5C
\mathchardef\measuredangle="0\msx@5D
\mathchardef\sphericalangle="0\msx@5E
\mathchardef\varpropto="3\msx@5F
\mathchardef\smallsmile="3\msx@60
\mathchardef\smallfrown="3\msx@61
\mathchardef\Subset="3\msx@62
\mathchardef\Supset="3\msx@63
\mathchardef\Cup="2\msx@64

\mathchardef\Cap="2\msx@65

\mathchardef\curlywedge="2\msx@66
\mathchardef\curlyvee="2\msx@67
\mathchardef\leftthreetimes="2\msx@68
\mathchardef\rightthreetimes="2\msx@69
\mathchardef\subseteqq="3\msx@6A
\mathchardef\supseteqq="3\msx@6B
\mathchardef\bumpeq="3\msx@6C
\mathchardef\Bumpeq="3\msx@6D
\mathchardef\lll="3\msx@6E

\mathchardef\ggg="3\msx@6F

\mathchardef\circledS="0\msx@73
\mathchardef\pitchfork="3\msx@74
\mathchardef\dotplus="2\msx@75
\mathchardef\backsim="3\msx@76
\mathchardef\backsimeq="3\msx@77
\mathchardef\complement="0\msx@7B
\mathchardef\intercal="2\msx@7C
\mathchardef\circledcirc="2\msx@7D
\mathchardef\circledast="2\msx@7E
\mathchardef\circleddash="2\msx@7F
\def\ulcorner{\delimiter"4\msx@70\msx@70 }
\def\urcorner{\delimiter"5\msx@71\msx@71 }
\def\llcorner{\delimiter"4\msx@78\msx@78 }
\def\lrcorner{\delimiter"5\msx@79\msx@79 }
\def\yen{\mathhexbox\msx@55 }
\def\checkmark{\mathhexbox\msx@58 }
\def\circledR{\mathhexbox\msx@72 }
\def\maltese{\mathhexbox\msx@7A }
\mathchardef\lvertneqq="3\msy@00
\mathchardef\gvertneqq="3\msy@01
\mathchardef\nleq="3\msy@02
\mathchardef\ngeq="3\msy@03
\mathchardef\nless="3\msy@04
\mathchardef\ngtr="3\msy@05
\mathchardef\nprec="3\msy@06
\mathchardef\nsucc="3\msy@07
\mathchardef\lneqq="3\msy@08
\mathchardef\gneqq="3\msy@09
\mathchardef\nleqslant="3\msy@0A
\mathchardef\ngeqslant="3\msy@0B
\mathchardef\lneq="3\msy@0C
\mathchardef\gneq="3\msy@0D
\mathchardef\npreceq="3\msy@0E
\mathchardef\nsucceq="3\msy@0F
\mathchardef\precnsim="3\msy@10
\mathchardef\succnsim="3\msy@11
\mathchardef\lnsim="3\msy@12
\mathchardef\gnsim="3\msy@13
\mathchardef\nleqq="3\msy@14
\mathchardef\ngeqq="3\msy@15
\mathchardef\precneqq="3\msy@16
\mathchardef\succneqq="3\msy@17
\mathchardef\precnapprox="3\msy@18
\mathchardef\succnapprox="3\msy@19
\mathchardef\lnapprox="3\msy@1A
\mathchardef\gnapprox="3\msy@1B
\mathchardef\nsim="3\msy@1C
\mathchardef\napprox="3\msy@1D
\mathchardef\varsubsetneq="3\msy@20
\mathchardef\varsupsetneq="3\msy@21
\mathchardef\nsubseteqq="3\msy@22
\mathchardef\nsupseteqq="3\msy@23
\mathchardef\subsetneqq="3\msy@24
\mathchardef\supsetneqq="3\msy@25
\mathchardef\varsubsetneqq="3\msy@26
\mathchardef\varsupsetneqq="3\msy@27
\mathchardef\subsetneq="3\msy@28
\mathchardef\supsetneq="3\msy@29
\mathchardef\nsubseteq="3\msy@2A
\mathchardef\nsupseteq="3\msy@2B
\mathchardef\nparallel="3\msy@2C
\mathchardef\nmid="3\msy@2D
\mathchardef\nshortmid="3\msy@2E
\mathchardef\nshortparallel="3\msy@2F
\mathchardef\nvdash="3\msy@30
\mathchardef\nVdash="3\msy@31
\mathchardef\nvDash="3\msy@32
\mathchardef\nVDash="3\msy@33
\mathchardef\ntrianglerighteq="3\msy@34
\mathchardef\ntrianglelefteq="3\msy@35
\mathchardef\ntriangleleft="3\msy@36
\mathchardef\ntriangleright="3\msy@37
\mathchardef\nleftarrow="3\msy@38
\mathchardef\nrightarrow="3\msy@39
\mathchardef\nLeftarrow="3\msy@3A
\mathchardef\nRightarrow="3\msy@3B
\mathchardef\nLeftrightarrow="3\msy@3C
\mathchardef\nleftrightarrow="3\msy@3D
\mathchardef\divideontimes="2\msy@3E
\mathchardef\varnothing="0\msy@3F
\mathchardef\nexists="0\msy@40
\mathchardef\mho="0\msy@66
\mathchardef\thorn="0\msy@67
\mathchardef\beth="0\msy@69
\mathchardef\gimel="0\msy@6A
\mathchardef\daleth="0\msy@6B
\mathchardef\lessdot="3\msy@6C
\mathchardef\gtrdot="3\msy@6D
\mathchardef\ltimes="2\msy@6E
\mathchardef\rtimes="2\msy@6F
\mathchardef\shortmid="3\msy@70
\mathchardef\shortparallel="3\msy@71
\mathchardef\smallsetminus="2\msy@72
\mathchardef\thicksim="3\msy@73
\mathchardef\thickapprox="3\msy@74
\mathchardef\approxeq="3\msy@75
\mathchardef\succapprox="3\msy@76
\mathchardef\precapprox="3\msy@77
\mathchardef\curvearrowleft="3\msy@78
\mathchardef\curvearrowright="3\msy@79
\mathchardef\digamma="0\msy@7A
\mathchardef\varkappa="0\msy@7B
\mathchardef\hslash="0\msy@7D
\mathchardef\hbar="0\msy@7E
\mathchardef\backepsilon="3\msy@7F
\def\Bbb{\ifmmode\let\next\Bbb@\else
 \def\next{\errmessage{Use \string\Bbb\space only in math mode}}\fi\next}
\def\Bbb@#1{{\Bbb@@{#1}}}
\def\Bbb@@#1{\fam\msyfam#1}

\catcode`\@=12
\font\teneusmf=eufm10 scaled 1200
\font\seveneusmf=eufm8
\font\fiveeusmf=eufm6
\newfam\eusmffam
\textfont\eusmffam=\teneusmf
\scriptfont\eusmffam=\seveneusmf
\scriptscriptfont\eusmffam=\fiveeusmf

\font\teneusm=eusm10 scaled 1200
\font\seveneusm=eusm8
\font\fiveeusm=eusm6
\newfam\eusmfam
\textfont\eusmfam=\teneusm
\scriptfont\eusmfam=\seveneusm
\scriptscriptfont\eusmfam=\fiveeusm

\font\teneusmc=cmsy10 scaled 1200
\font\seveneusmc=cmsy8
\font\fiveeusmc=cmsy6
\newfam\eusmcfam
\textfont\eusmcfam=\teneusmc
\scriptfont\eusmcfam=\seveneusmc
\scriptscriptfont\eusmcfam=\fiveeusmc

\newcommand{\titletextCERN}
{
Prospects for a
Measurement of \boldmath $\alpha_s$ \unboldmath\\
via Scaling Violations of Fragmentation Functions\\
in Deeply Inelastic Scattering
}

\newcommand{\titletextPL}
{
Prospects for a
Measurement of \boldmath $\alpha_s$ \unboldmath\\
via Scaling Violations of Fragmentation Functions\\
in Deeply Inelastic Scattering\\
}

\newcommand{\abstracttext}
{
The prospects for a determination of the strong coupling constant $\alpha_s$
via scaling violations of fragmentation functions in deeply inelastic scattering
are studied. The statistical error in the case of an integrated
luminosity of $250\,\mbox{pb}^{-1}$, and the theoretical errors due to
the various parton density parametrizations and to the factorization
scale dependence are estimated.
}

\newlength{\dinwidth}                       
\newlength{\dinmargin}                      
\ifnum\cerntp=0
\fi
\def\lsim{\mathrel{\rlap{\lower4pt\hbox{\hskip1pt$\sim$}}
    \raise1pt\hbox{$<$}}}                
\def\gsim{\mathrel{\rlap{\lower4pt\hbox{\hskip1pt$\sim$}}
    \raise1pt\hbox{$>$}}}                
\begin{document}


\ifnum\cerntp=1

\thispagestyle{empty}

\renewcommand{\thefootnote}{\fnsymbol{footnote}}
\setcounter{footnote}{0}

\begin{flushright}
{
\unitlength 1mm
\begin{picture}(10,10)
\put(0,0){CERN--TH/\cernnr}
\put(0,-5){hep-ph/9610287}
\end{picture}
\rule{2cm}{0mm}
}
\end{flushright}
 
\vspace{1.5cm}

\begin{center}

{\Large\bf \titletextCERN$\!\!$\footnote[3]{\it Based on a contribution to the 
workshop ``Future Physics at HERA'', DESY, Hamburg, May 1996.
}\\}


\vspace{1.5cm}

{\bf Dirk~Graudenz}\footnote[1]{\em Electronic
mail address: Dirk.Graudenz\char64{}cern.ch}\footnote[2]{\em
WWW URL: http://wwwcn.cern.ch/$\sim$graudenz/index.html}


{\it Theoretical Physics Division, CERN\\
1211 Geneva 23, Switzerland}

\end{center}

\vspace{1.0cm}

\begin{center}
{\bf Abstract}
\end{center}

\hspace{4mm}
\abstracttext

\vfill
\noindent
CERN--TH/\cernnr\\
\vdate

\clearpage
\setcounter{page}{1}

\fi
 

\renewcommand{\thefootnote}{\fnsymbol{footnote}}
\setcounter{footnote}{0}

\vspace*{1cm}
\begin{center}  \begin{Large} \begin{bf}
\titletextPL
  \end{bf}  \end{Large}
  \vspace*{5mm}
  \begin{large}
Dirk~Graudenz$\:\!^a\;\!$\footnote[1]{\em Electronic
mail address: Dirk.Graudenz\char64{}cern.ch}$\:\!$\footnote[2]{\em
WWW URL: http://wwwcn.cern.ch/$\sim$graudenz/index.html}\\
  \end{large}
\end{center}
$^a$ Theoretical Physics Division, CERN, 1211 Geneva 23, Switzerland\\
%
\begin{quotation}
\noindent
{\bf Abstract:}
\abstracttext
\end{quotation}

\renewcommand{\thefootnote}{\arabic{footnote}}
\setcounter{footnote}{0}


\section{Introduction}
The strong coupling constant $\alpha_s$ has been measured at HERA by means of 
the (2+1) jet rate~\cite{1}. 
This particular process has the advantage that
$\alpha_s$ is, up to higher order corrections, directly proportional to
the ratio of the measured cross-sections. Another route to a determination of
$\alpha_s$ is given by scaling violations of phenomenological distribution
functions. Perturbative QCD predicts the scale evolution of these quantities
by means of renormalization group equations \cite{2,3}. 
As a consequence, again up to 
higher order corrections and resummation effects, 
the experimentally measurable quantities
(the structure functions, and the fragmentation functions,
which depend on factorization scale and scheme)
are (symbolically) of the form
$A+B\,\alpha_s\ln\mu^2$. Here $\mu$ is the factorization scale, which 
is to be identified,
in deeply
inelastic scattering, 
with a scale of the order of 
the photon virtuality $Q$,
for lack of other hard scales related to the 
leading-order process. The distribution functions contain a 
$\mu$-independent
term $A$, and the $\alpha_s$~dependence is only logarithmic in the 
factorization scale. The scaling violations are therefore expected 
to be small, and
will require large luminosity to be statistically significant. 
An $\alpha_s$~determination via scaling violations
has the advantage
that,
in principle, no explicit model assumptions
such as specific fragmentation models go into the measurement. In 
the case of scaling violations of structure functions\footnote{See, for 
example, the contributions of the working group on 
structure functions.}
a completely inclusive quantity is measured, and the theoretical
basis, namely the operator product expansion, is very transparent and can be 
derived rigorously from light cone dominance.
In the case of one-particle-inclusive processes, where the
operator product expansion is not available, the factorization
theorem of perturbative QCD (see, for example, Ref.~\cite{4} and references 
therein) allows the separation of the hard scattering process from the 
non-perturbative fragmentation process. The one-particle-inclusive 
cross-section is a convolution of a mass-factorized parton-level scattering
cross-section, a parton density and a fragmentation function:
$\sigma=\sigma_{\mbox{\scriptsize hard}}\otimes f \otimes D$. A possible 
strategy for an $\alpha_s$~measurement at HERA is to perform a 
combined multiparameter
fit of fragmentation functions and of the strong coupling constant to
the $x_F$-distribution $\rho\left(x_F\right)=\left(\dd \sigma/\dd x_F\right) 
/ \sigma_{\mbox{\scriptsize tot}}$
(or to any other distribution sensitive to 
the fragmentation functions) of charged hadrons
at two different scales~$Q$.
Here the variable $x_F$ is defined to be $2h_L/W$, where $h_L$ is the
longitudinal momentum fraction of the observed charged hadron 
in the direction of the virtual photon in the hadronic centre-of-mass 
frame\footnote{It might be possible to reduce the dependence on the parton 
densities and on the not yet well understood physics of the forward direction
by performing an analysis in the Breit frame. I thank N.~Brook and T.~Doyle
for remarks concerning this issue.},
and $W$ is the total hadronic final state energy. In leading order, $x_F$
is the momentum fraction of the final state 
current quark carried by the observed 
hadron. The total cross-section is denoted by 
$\sigma_{\mbox{\scriptsize tot}}$. The strong coupling constant enters the
expression for $\rho$ in three places: (a) as an expansion parameter in the
next-to-leading order expression for $\sigma_{\mbox{\scriptsize hard}}$, 
and in the renormalization group equations of (b) the fragmentation functions
and (c) the parton densities. The parton densities are an input to the analysis.
Since they are obtained by a global fit, where a specific value of $\alpha_s$
is used, it is necessary to include this dependence as well as 
the variation due to the different parametrizations into the systematic
error\footnote{Alternatively, parton density parametrizations
with varying values of the strong coupling constant \cite{5}
can be employed.}.
The next-to-leading-order one-particle-inclusive cross-section has been 
calculated in Ref.~\cite{6}. For our study, we use a recent recalculation
and numerical implementation described in Ref.~\cite{7}.
A comparison of the theoretical $x_F$-distribution with experimental data
from the H1 and ZEUS Collaborations \cite{8}
has been done in Ref.~\cite{9}. It turns out that the theoretical 
description of the experimental data is quite satisfactory. The next-to-leading 
order result is always within one standard deviation of the experimental
data points except for those at very large~$x_F$, where the currently available
fragmentation function parametrizations are not well constrained 
by \epem{} data.

In the next section, we describe the estimate of the various errors\footnote{
We do not consider experimental systematic errors.
} 
of the
value of $\alpha_s$. We also discuss the sensitivity to the strong coupling
constant of various ranges in~$x_F$. The paper closes with a short summary and
conclusions.

\section{Error Estimates}
To get a quantitative estimate of the dependence of the scale evolution 
of fragmentation functions on the employed value of $\alpha_s$, 
we fix the
fragmentation functions at a scale of $\mu_0=2\,\GeV$ as the leading-order
parametrization of Ref.~\cite{10}.
We then evolve this input with two 
different values 
for $\Lambda_{\mbox{\scriptsize QCD}}^{(4)}$
of $0.1\,\GeV$ \{a\} and $0.2\,\GeV$ \{b\}.
The corresponding 
$x_F$-distributions
$\rho^{\{a\}}$ and $\rho^{\{b\}}$ 
are determined for these two sets of fragmentation functions.
We now assume that the $x_F$-distributions $\rho$ are measured in two different
bins $i$, $j$ of the factorization scale~$Q$. The ratios 
$\lambda^{\{ij\}}=\rho^{\{i\}}/\rho^{\{j\}}$
for an arbitrary coupling constant 
$\alpha_s$ (taken at the mass of the $Z$~boson)
are expanded in a power series in 
$\alpha_s$, where only the linear term is kept: 
\beqm{ratio}
\lambda^{\{ij\}}=\lambda^{\{aij\}}
+\frac{\lambda^{\{bij\}}-\lambda^{\{aij\}}}
      {\alpha_s^{\{b\}}-\alpha_s^{\{a\}}}\,
\left(\alpha_s-\alpha_s^{\{a\}}\right).
\eeq
Based on this formula, a quantitative estimate of the statistical error of
$\alpha_s$ is possible. Moreover, by varying $\lambda^{\{ij\}}$, for
example by using various parton density distributions or by modifying the
factorization scale, the impact of systematic effects can be studied.

\begin{figure}[htb] \unitlength 1mm
\begin{center}
\dgpicture{159}{53}

\put(40,-5){\epsfigdg{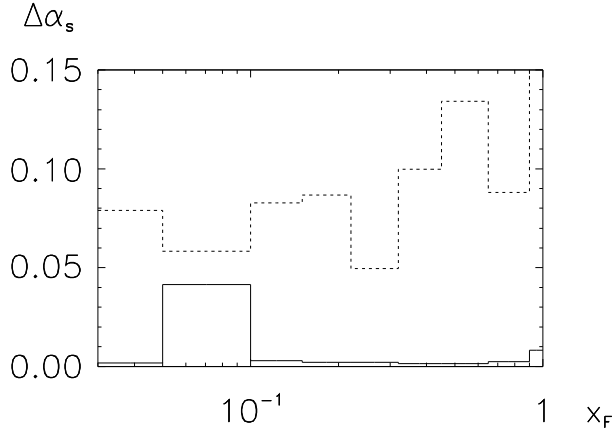}{width=80mm}} 

\end{picture}
\end{center}
\caption[]
{\labelmm{delas} {\it 
Statistical errors of $\alpha_s$ for the individual $x_F$-bins,
for the $Q$-bins {\rm \{1\}}, {\rm \{2\}} 
\mbox{\rm[\fullline]} and {\rm \{2\}}, {\rm \{3\}}
\mbox{\rm[\dashline]}.
}}
\end{figure}

To be more specific, we assume the phase space cuts and $x_F$-bins
of the ZEUS analysis,
except for the range in the photon virtuality $Q$, where 
we consider three bins:
$[3.16,12.6]\,\GeV$ \{1\}, 
$[12.6,100]\,\GeV$ \{2\} and
$[100,150]\,\GeV$ \{3\}. 
To obtain explicit numerical values, we use the CTEQ 3L
parametrization \cite{11} for the
parton densities 
(for simplicity, we work in leading order).
The integrated luminosity is assumed to be $250\,\mbox{pb}^{-1}$.
Under the assumption of Gaussian statistical errors we arrive at 
a statistical error 
of $\alpha_s(M_Z^2)$ of
$\Delta\alpha_s^{\mbox{\scriptsize stat}}=$ $\pm 0.0007$ 
for an analysis based on bins~\{1\} and~\{2\} and
of $\pm 0.027$
for an analysis based on bins~\{2\} and~\{3\}.
The individual errors of $\alpha_s$ for the various $x_F$-bins are shown
in Fig.~\ref{delas}. The large error around $x_F\sim 0.08$ for the large-$Q$
bins comes from the fact that the evolution of the fragmentation functions 
around this value of the momentum fraction is quite small (for smaller 
values, the fragmentation functions increase with increasing factorization 
scale, and for larger values, they decrease). The sensitivity of the 
cross-section to a variation of $\alpha_s$ is largest at large $x_F$, but
this region also suffers from small statistics of the data sample. It turns 
out that the full $x_F$-range is about equally important.

As briefly mentioned already in the introduction, an input parton density
has to be chosen. 
To estimate the size of this effect, we determine the spread of the results
for $\alpha_s(M_Z^2)$ 
depending on the next-to-leading-order parton densities 
from Refs.~\cite{11,12}. For the bins~\{1\} and~\{2\}, 
the spread is
$\Delta\alpha_s^{\mbox{\scriptsize PDF}}=$
$\pm 0.017$, and for the bins~\{2\} and~\{3\}, it is
$\pm 0.005$.
Future global fits of parton densities including improved HERA data
should reduce this systematic uncertainty.

Perturbative QCD allows for some freedom in the choice of the 
factorization scale~$\mu$ of the fragmentation functions $D(z,\mu^2)$.
This brings out the inherent uncertainty in the theoretical prediction, 
and can be interpreted as an effect of unknown higher order contributions.
To obtain an estimate of this uncertainty, the ratios~$\lambda$ 
are determined for the three choices $Q/2$, $Q$ and $2Q$ of this
scale. The change of cross-section has for consequence a variation in 
the extracted $\alpha_s(M_Z^2)$ value of 
$\Delta\alpha_s^{\mbox{\scriptsize scale}}=$ $\pm 0.013$ and 
$\pm 0.011$ for the
combinations of the bins \{1\},~\{2\} and \{2\},~\{3\}, respectively.

\ifnum\opus=1
\begin{figure}[htb] \unitlength 1mm
\begin{center}
\dgpicture{159}{70}

\put(60,0){\epsfigdg{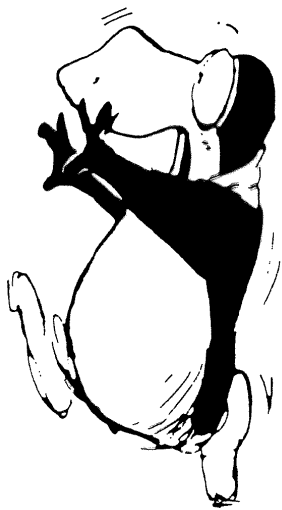}{width=40mm}} 

\end{picture}
\end{center}
\caption[]
{\labelmm{rc} {\it 
The running strong coupling constant, as recently observed in high energy
collider experiments.
}}
\end{figure}
\fi

\section{Summary and Conclusions}
We have studied the prospects of a measurement of the strong coupling constant
in deeply inelastic scattering at HERA by means of scaling violations of
fragmentation functions. 
The combinations of the obtained 
values for $\Delta\alpha_s^{\mbox{\scriptsize stat}}$, 
$\Delta\alpha_s^{\mbox{\scriptsize PDF}}$
and 
$\Delta\alpha_s^{\mbox{\scriptsize scale}}$
are large compared with the present error $\Delta\alpha_s=0.006$ of the
world average. It is therefore likely that a measurement of this kind
will not be competitive, concerning the size of the error.
Nevertheless, 
it is worth doing 
as an independent quantitative test of QCD and, more important, 
because it 
complements the other (potential) HERA measurements based 
on (2+1) jet rates and scaling violations of structure 
functions.
\ifnum\opus=1
The prospects for the observation of the running of $\alpha_s$ (Fig.~\ref{rc}) 
should also be studied in some detail.
\fi

\medskip
\medskip
\medskip
\noindent
{\Large \bf Acknowledgements}

\medskip
\noindent
I wish to thank Ch.~Berger, N.~Brook, T.~Doyle, M.~Kuhlen and N.~Pavel for 
discussions, H.~Spiesberger for comments on the manuscript, 
and J.~Binnewies for clarifying remarks concerning
the parametrizations of Ref.~\cite{10}.
This work was supported in part by a Habilitandenstipendium of the
Deutsche Forschungsgemeinschaft.


\newcommand{\scs}{\rm}
\newcommand{\bibitema}[1]{\bibitem[#1]{#1}}
\newcommand{\bibbeginshort}{\begin{thebibliography}{XX}}
\newcommand{\bibbeginlong}{\begin{thebibliography}{XXX\,1988\,\,}}

\newcommand{\bibitemb}[4]{\begin{sloppy}
{\bibitem[#1]{#1} {#2}, {#4.}}
\end{sloppy}}

\newcommand{\nbibitemb}[4]{}

\bibbeginshort 

   \nbibitemb{H195a}
      {H1 Collaboration, T.~Ahmed et al.}
      {}
      {Phys.\ Lett.\ \underline{B346} (1995) 415}

   \nbibitemb{BD96} 
      {N.~Brook and T.~Doyle}
      {}
      {private communication}

   \nbibitemb{CFP80}
      {G.~Curci, W.~Furmanski and R.~Petronzio}
      {Evolution of Parton Densities Beyond Leading Order}
      {Nucl.\ Phys.\ \underline{B175} (1980) 27}

   \nbibitemb{NW94}
      {P.~Nason and B.R.~Webber}
      {}
      {Nucl.\ Phys.\ \underline{B421} (1994) 473}

   \nbibitemb{H195b} 
      {H1 Collaboration, S.~Aid et al.,}
      {} 
      {preprint DESY 95-086 (1995)}

%
\bibitema{1}
      T.~Ahmed et al.\ (H1 Collaboration),
      Phys.\ Lett.\ \underline{B346} (1995) 415;
      M.~Derrick et al.\ (ZEUS Collaboration),
      Phys.\ Lett.\ \underline{B363} (1995) 201.

\bibitemb{2}
      {G.~Altarelli and G.~Parisi}
      {Asymptotic Freedom in Parton Language}
      {Nucl.\ Phys.\ \underline{B126} (1977) 298}

\bibitemb{3} 
      {R.~Baier and K.~Fey}
      {Finite Corrections to Quark Fragmentation Functions in  
       Perturbative QCD}
      {Z.~Phys.\ \underline{C2} (1979) 339}

\bibitemb{4} 
      {J.C.~Collins, D.E.~Soper and G.~Sterman}
      {Factorization of Hard Processes in QCD}
      {in: {\it Perturbative Quantum Chromodynamics,} ed.\ A.H.~Mueller
       (World Scientific, Singapore, 1989)}

\bibitema{5} 
      A.~Vogt, Phys.\ Lett.\ \underline{B354} (1995) 145;
      A.D.~Martin, W.J.~Stirling and R.G.~Roberts, 
      Rutherford Appleton Lab.\ preprint RAL-TR-95-013 (June 1995).

\bibitemb{6}
      {G.~Altarelli, R.K.~Ellis, G.~Martinelli and S.Y.~Pi}
      {Processes Involving Fragmentation Functions Beyond the 
       Leading Order in QCD}
      {Nucl.\ Phys.\ \underline{B160} (1979) 301}

\bibitemb{7}
       {D.~Graudenz}
       {}
       {{\it Heavy-Quark Production in the Target Fragmentation Region,} 
        preprint CERN--TH/96--52, in preparation} 

\bibitema{8}
      I.~Abt et al.\ (H1 Collaboration)
      Z.~Phys.\ \underline{C63} (1994) 377;
      M.~Derrick et al.\ (ZEUS Collaboration),
      Z.~Phys.\ \underline{C70} (1996) 1.

\bibitema{9} 
      D.~Graudenz, 
      {\it Charged-Meson Production and Scaling Violations
       of Fragmentation Functions in Deeply Inelastic Scattering 
       at HERA,} preprint CERN--TH/96--155 (June 1996).

\bibitema{10}
      J.~Binnewies, B.A.~Kniehl and G.~Kramer,
      Z.~Phys.\ \underline{C65} (1995) 471;
      Phys.\ Rev.\ \underline{D52} (1995) 4947.

\bibitemb{11}
      {CTEQ Collaboration}
      {Global QCD Analysis and the CTEQ Parton Distributions}
      {Phys.\ Rev.\ \underline{D51} (1995) 4763}

\bibitema{12}
       M.~Gl\"{u}ck, E.~Reya and A.~Vogt,
       Z.~Phys.\ \underline{C67} (1995) 433; 
       A.D.~Martin, R.G.~Roberts and W.J.~Stirling,
       Phys.\ Rev.\ \underline{D51} (1995) 4756.

\end{thebibliography}

\end{document}

===============================================================================